\documentclass[preprint2,trackchanges,lineno]{aastex6}

\newcommand{\msun}{M_{\odot}}

\fullcollaborationName{The Friends of AASTeX Collaboration}

\begin{document}


\title{Super-Knee Cosmic Rays from Galactic Neutron Star Merger Remnants}


\author{Shigeo S. Kimura\altaffilmark{1,2,3}, Kohta Murase\altaffilmark{1,2,3,4}, and Peter M\'{e}sz\'{a}ros\altaffilmark{1,2,3}}
\altaffiltext{1}{Department of Physics, Pennsylvania State University, University Park, Pennsylvania 16802, USA}
\altaffiltext{2}{Department of Astronomy \& Astrophysics, Pennsylvania State UNiversity, University Park, Pennsylvania 16802, USA}
\altaffiltext{3}{Center for Particle and Gravitational Astrophysics, Pennsylvania State University, University Park, Pennsylvania 16802, USA}
\altaffiltext{4}{Center for Gravitational Physics, Yukawa Institute for Theoretical Physics, Kyoto, Kyoto 606-8502, Japan}


\begin{abstract}
The detection of gravitational waves and electromagnetic counterparts from a binary neutron star (BNS) merger confirmed that it is accompanied by the launch of fast merger ejecta. Analogous to supernova remnants, forward shocks formed by the interaction of the ejecta with interstellar material will produce high-energy cosmic rays. 
We investigate the possibility that Galactic neutron star merger remnants (NSMRs) significantly contribute to the observed cosmic rays in the energy range between the knee and the ankle. Using typical parameters obtained by modeling of GW170817, we find that NSMRs can accelerate iron nuclei up to $\sim500$~PeV. 
We calculate the cosmic-ray spectrum and composition observed on Earth, and show that the Galactic NSMR scenario can account for the experimental cosmic-ray data in the $20-1000$~PeV range.
Our model can naturally explain the hardening feature around 20~PeV for the total cosmic-ray spectrum, which has been observed by the Telescope Array Low Energy extension and the IceTop air shower array.  
\end{abstract}

\keywords{cosmic rays --- acceleration of particles --- astroparticle physics  ---  ISM: supernova remnants --- gravitational waves --- stars: neutron}



\section{Introduction} \label{sec:intro}

The origin of the diffuse cosmic-ray (CR) flux observed on Earth is one of the greatest mysteries in high-energy astrophysics. Direct-detection and air-shower experiments have revealed that the spectrum of CRs is described by a power-law function, $\Phi\propto E^{-\gamma}$ with a few break points~\citep[see, e.g.,][for reviews]{NW00a,KO11a}. The first break appears around $\sim (3-4)\times10^{15}$ eV (``knee''), where the spectral index changes from $\gamma\approx2.7$ to $\gamma\approx3.0$. A second knee appears around $10^{17}$ eV, above which the spectrum is softened to $\gamma\approx3.2-3.3$. The third break, called the ankle, hardens the spectrum to $\gamma\approx2.6$ around $(3-5)\times10^{18}$ eV. The final break is located around $6\times10^{19}$ eV, which is consistent with the cutoff caused by interactions with the cosmic microwave background and extragalactic background light.

Galactic supernova remnants (SNRs) are believed to be responsible for CRs below the knee (see \citealt{Dru83a,Hil05a,Bel13a} for reviews). However, the recent studies of historical SNRs have indicated that the maximum energy is lower than the knee energy~\citep[see, e.g.,][]{Aha13a,MAGIC17a,HESS18a}. 
Different possibilities have been discussed to explain the CRs beyond the knee energy. Those include core-collapse supernovae with dense circumstellar media~\citep[e.g.,][]{Sve03a,MTO14a,ZP16a}, the Galactic Center \citep[Sgr A*; e.g.,][]{HESS17a,FMK17a,AYO18a,GRK18a}, and highly spinning black holes created by binary black-hole mergers \citep{IMT17a}.
On the other hand, the CRs above the ankle, which are often called ``ultrahigh-energy cosmic rays (UHECRs)'', should originate from extragalactic sources, such as active galactic nuclei \citep[e.g.,][]{Tak90a,ps92,BGG06a,MDT12a,FM18a,KMZ18a,RFG18a}, gamma-ray bursts \citep[e.g.,][]{Wax95a,Vie96a,MIN08a,GAM15a,AM16a,BBFF18a,ZMK18a}, and magnetars \citep[e.g.,][]{Aro03a,MMZ09a,FKM14a}. 
It seems that another component (that is often called the ``B'' component) is needed to fill the gap between these two components~\citep{Hil05a,Gai12a,GST13a}. 
This second component should accelerate the CRs up to higher energies than the ordinary SNRs. This requires the combination of a higher shock velocity, a larger size, and a stronger magnetic field. The candidate sources include Galactic supernovae with dense winds~\citep[e.g.,][]{Sve03a,PZS10a,MTO14a,ZP16a}, Galactic winds~\citep{JM87a,VZ04a,TRV16a,BZC17a,MF18a}, Galactic newborn pulsars~\citep{FKO13a}, Galactic gamma-ray bursts~\citep{LE93a,WDA04a,CKN10a}, trans-relativistic supernovae~\citep{WRM07a,BKM08a}, and galaxy clusters~\citep{min08,FM18a}. 

In 2017, gravitational waves from a merger of a binary neutron star (BNS) was detected, followed by the electromagnetic counterparts~\citep{LIGO17c,LIGO17d}. 
The slowly brightening afterglow emission~\citep[e.g.,][]{LLL18a,MNH18a,GMRT18a,CHANDRA18a,TPR18a} imply the existence of relativistic jets in this system \citep[e.g.,][]{LPM17a,MAX18a}, 
and hadronic production of high-energy neutrinos are also discussed~\citep{KMM17b,BHW18a,KMB18a}. 
Besides, the UV/optical/IR counterparts powered by radioactive nuclei (kilonova/macronova) enabled us to confirm that BNS mergers generate fast and massive outflows \citep{ALOP17a,LCO17a,CBV17a,TROS17a,DPS17Sa,ECK17a,AST17a,KNS17a,MASTER17a,GRAWITA17a,DES17a,JGEM17a,DLT4017a}. 
The merger ejeca should accelerate particles at shocks formed by interaction with the interstellar material (ISM), forming a neutron star merger remnant (NSMR) analogous to an SNR.  
The leptonic afterglow emission of NSMRs has been intensively studied~\citep[e.g.,][]{NP11a,TKI14a,HNH16a}.
On the other hand, the hadronic CR production was not studied in detail. \citet{TKI14a} discussed the possibility that NSMRs contribute to the CRs around the ankle, without quantitative comparisons to the experimental results. 
In this paper, we study the possibility that NSMRs in the Milky Way significantly contribute to the CR flux beyond the knee. 
The paper is organized as follows. In Section \ref{sec:CRinNSMRs}, we discuss CR production at typical NSMRs through the estimation of physical quantities and the maximum CR energy. In Section \ref{sec:CRonEarth}, we approximately calculate the CR spectrum and composition on Earth and compare our results to the experimental data. We discuss the related issues in Section \ref{sec:discussion} and make conclusions in Section \ref{sec:summary}.
We use notations $Q_x = Q/10^x$ in CGS units otherwise noted.

\section{CR production at Merger Remnants}\label{sec:CRinNSMRs}
\subsection{Physical quantities}\label{sec:quantities}
Theoretical modeling of GW170817 has revealed that the velocity and mass of the merger ejecta are $V_{\rm ini}\sim 0.1c-0.3c$ and $M_{\rm ej}\sim 0.01\rm\,\msun-0.05\rm\,\msun$, respectively \citep[e.g.,][]{KMB17a,MRK17a,RSF17a,SCJ17a,JGEM17b,WOK17a,MIK18a}, which leads to the energy of the ejecta, $\sim 10^{50}\rm\,erg\,-\,3\times10^{51}$ erg.
The ejecta can consist of multiple components: a fast-light component radiating the early UV/blue photons and a slow-heavy component emitting the red/IR photons later. Since the most energetic component is likely to dominate over the other components, we approximate the ejecta as a single component. 
 Initially, the merger ejecta freely expands into the ISM with $V_{\rm ej}\approx V_{\rm ini}$. 
After they sweep up the ISM with a mass of $\sim M_{\rm ej}$, they start to be decelerated. The Sedov time and radius are estimated to be
\begin{equation}
 R_{\rm dec}\approx \left(\frac{3M_{\rm ej}}{4\pi \mu m_p n_{\rm ISM}}\right)^{1/3} 
\simeq 3.9\times10^{18} M_{\rm ej,-1.5}^{1/3} n_{\rm ISM,-1}^{-1/3} \rm\,cm,
\end{equation}
\begin{equation}
 t_{\rm dec}\approx \frac{R_{\rm dec} }{V_{\rm ini}}
\simeq 5.2\times 10^{8} M_{\rm ej,-1.5}^{1/3} n_{\rm ISM,-1}^{-1/3} V_{\rm ini,-0.6}^{-1}\rm\,s,
\end{equation}
where $\mu\simeq 1.4$ is the mean molecular weight for ISM, $m_p$ is the proton mass, $n_{\rm ISM}$ is the mean number density of ISM. Here, we use $M_{\rm ej,-1.5}=M_{\rm ej}/(0.03\rm\,\msun)$ and $V_{\rm ini,-0.6}=V_{\rm ini}/(0.25c)$. For $t>t_{\rm dec}$, the time evolution of the ejecta radius and velocity are given by the Sedov-Taylor solution: $R_{\rm ej}\propto t^{2/5}$ and $V_{\rm ej}\propto t^{-3/5}$, respectively.

At the forward shock of the NSMR, CRs are likely to be produced, and CR driven instabilities can amplify the magnetic field around the shock \citep[e.g.,][]{Bel04a}. 
Details of the magnetic amplification at collisionless shocks are currently not well understood, so we use a simple parameterization using a constant $\epsilon_B$ parameter:
\begin{equation}
 B \approx \sqrt{4\pi  \epsilon_B \mu m_p n_{\rm ISM} V_{\rm ej}^2}
  \simeq 0.41 n_{\rm ISM,-1}^{1/2}V_{\rm ini,-0.6}\epsilon_{B,-3}^{1/2}  \rm\, mG,
\end{equation}
where $\epsilon_B$ is the ratio of the magnetic field pressure to the ram-pressure, and use $V_{\rm ej}\approx V_{\rm ini}$ at the second equation. We assume that the upstream magnetic field is also amplified to the similar value for simplicity.

\subsection{Maximum energy}\label{sec:emax}
The maximum energy of the accelerated particles of species $i$, $E_{i,\rm max}$, is determined by the balance between the acceleration and either age of the NSMR or diffusive escape time.
For a quasi-parallel shock, the particle acceleration time is estimated to be \citep[e.g.,][]{Dru83a,BE87a}
\begin{equation}
t_{\rm acc}\approx \frac{20}{3} \frac{r_{L,i}}{c} \left(\frac{V_{\rm ej}}{c}\right)^{-2},
\end{equation}
where $r_{L,i}=E/(Z_ieB)$ is the Larmor radius of the particle species $i$ ($Z_i$ is the charge of the particle). 
Some fraction of the forward shock is quasi-perpendicular to the magnetic field, where the CR spectrum can be steeper than that at the quasi-parallel shocks. However, even if we take this effect into account, the averaged spectrum is expected to be similar to that with the parallel shock for the parameter range of our interest \citep{BSR11a}.

Equating the acceleration time to the age of the NSMR, $t_{\rm age}\approx R_{\rm ej}/V_{\rm ej}$, we obtain
\begin{equation}
 E_{i,\rm max} \approx \frac{3Z_i e B R_{\rm ej} V_{\rm ej}}{20 c }.\label{eq:emax}
\end{equation}
The diffusive escape time from the NSMR is estimated to be $t_{\rm diff}\approx R_{\rm ej}^2/(2D_{\rm SNR})\sim R_{\rm ej}^2/(r_{L,i} c)$, where we use $D_{\rm SNR}\approx r_{L,i} c/3$. Then, equating the diffusion time to the acceleration time, we obtain the formula similar to equation (\ref{eq:emax}).
The maximum energy is proportional to $t$ during the free-expansion phase, while $E_{i,\rm max}\propto t^{-4/5}$ during the Sedov phase.
The maximum value of $E_{i,\rm max}$ is given at $t\sim t_{\rm dec}$, which is estimated to be:
\begin{equation}
 E_{i,\rm max} \simeq 1.8\times10^{16} M_{\rm ej,-1.5}^{1/3} n_{\rm ISM,-1}^{1/6}V_{\rm ini,-0.6}^{2} \epsilon_{B,-3}^{1/2}  Z_i \rm\, eV.
\end{equation}
Thus, the NSMRs can accelerate protons and iron nuclei to energies higher than the knee and the second knee, respectively. 

\subsection{Spectrum of escaping CRs}\label{sec:EscapeSpectrum}
We can observe only the CRs which escape from the NSMR. The escape condition can be represented as $t_{\rm diff} \lesssim t_{\rm age}$. The critical energy at which $t_{\rm diff}\sim t_{\rm age}$ is close to that given by equation (\ref{eq:emax}). Thus, low energy CRs are confined in the NSMR, and only the particles around $E_{i,\rm max}$ can escape to ISM. Although details of the escape process are not well known, we make a brief discussion based on \citet{OMY10a}.

If the maximum energy of CRs and/or the differential CR number density, $N_{E,\rm src}$, at the NSMR varies with time, the time integration of the escaping CRs can create a power-law spectrum. If we write $E_{i,\rm max}\propto t^{-\alpha}$ and $N_{E,\rm src}\propto E^{-s_{\rm inj}}t^{\beta}$, the index of the escape CR spectrum is represented as $s_{\rm esc}= s_{\rm inj} + \beta/\alpha$ (see Appendix of \citealt{OMY10a} for applicable parameter space).

If we assume a constant CR production efficiency, the CR density can be written as $N_{E,\rm src}\propto V_{\rm ej}^2 R_{\rm ej}^3$, leading to $N_{E,\rm src}\propto t^3$ and $N_{E,\rm src}\propto t^0$ for the free-expansion phase and the Sedov phase, respectively. Since the CR production rate is much higher during the Sedov phase, we can safely neglect the CR production during the free expansion phase.
Then, we obtain $\beta\approx 0$, resulting in $s_{\rm esc}\simeq s_{\rm inj}$. Hence, the spectrum of escaping CRs is the same with the CRs at the source.

\subsection{Composition at the source}\label{sec:composition}

To estimate the abundance ratio of the CRs, we use a model proposed by \citet{CYS17a}, in which the injection efficiency for a heavy element is higher than that for protons by a factor $(A_i/Z_i)^2$ ($A_i$ is the mass number) at the injection energy. The injection occurs in a non-relativistic regime, and the spectral softening appears at $ E\sim A_i m_p c^2$. Then, the energy density, $E^2 dN_i/dE$, of heavy element at this energy is further enhanced by another factor of $(A_i/Z_i)^{1/2}$. We assume that ISM consists of singly ionized plasma with the solar abundance ratio (see, e.g., \citealt{Lod03a}). Then, the resulting abundance ratio at the NSMR at a given energy is estimated to be $(f_p,\, f_{\rm He},\, f_{\rm C},\, f_{\rm O},\, f_{\rm Ne},\, f_{\rm Si},\, f_{\rm Fe}) \simeq (0.17,\,0.52,\,0.024,\,0.099,\,0.027,\,0.028,\,0.14)$. 
The other elements are negligible. This abundance ratio is different from the abundance ratio on Earth due to propagation effects (see Section \ref{sec:CRonEarth}). 
The CR nuclei should be fully ionized after they are accelerated to relativistic energy.

\section{CR spectrum and composition at Earth}\label{sec:CRonEarth}
\subsection{Event rate vs escape time}

For NSMRs to be the sources of Galactic CRs, the occurrence time should be shorter than the CR escape time from the Galaxy. The merger rate inside our Galaxy is estimated to be $\rho_{\rm MW}\sim \rho_{\rm mer}/n_{\rm MW}\sim 1.5\times 10^{-4}\,\rm yr^{-1}$, where $\rho_{\rm mer}\sim 1.5\times10^{-6}\,\rm Mpc^{-3}\, yr^{-1}$ is the local BNS merger rate obtained by the GW detection \citep{LIGO17c} and $n_{\rm MW}\sim 0.01\,\rm Mpc^{-3}$ is the number density of the Milky-Way--size galaxies. This merger rate is consistent with the abundance of r-process elements in the Milky Way \citep{HBP18a}, although the rate has still large uncertainty.

The escape time of CRs from the CR halo, $T_{\rm esc}$, is estimated by the abundance of radioactive nuclei \citep[e.g.,][]{YWM01a,HBB04a}. The value of $T_{\rm esc}$ depends on the details of the models. For example, leaky box models give $T_{\rm esc}\sim 15-100$ Myr at 1 GV \citep{YWM01a,Blu11a}, while diffusion halo  models present longer escape timescales: $T_{\rm esc}\sim 20-400$ Myr at 10 GV \citep{SMP07a,Blu11a,Lip14a,Yua18a}. We often assume that $T_{\rm esc}\propto \mathcal R^{-\delta}$, where $\mathcal R=E/(Z_i e)$ is the rigidity and $\delta\sim0.3-0.6$ \citep{SMP07a}. If $\delta\lesssim0.4$, $T_{\rm esc} > \rho_{\rm MW}^{-1}\sim 6.7\times10^3$ yr is always satisfied for the range of our interest ($\mathcal R\lesssim 10^{8}$ GV). Thus, NSMRs can supply CRs before they escape from the Galaxy as long as  $\delta\lesssim 0.4$. 
Note that this does not guarantee the homegeneity of the CRs inside the ISM. This should be examined by modeling with an inhomogeneous source distribution (see also Section \ref{sec:discussion}).

Note also that we use a constant value of $\delta$ for $\mathcal R \lesssim 10^8$ GV, which is not guaranteed from the experimental data. Theoretically, the resonant scattering by Kolmogorov turbulence results in $\delta\simeq1/3$, which requires the Lamor radius of CRs to be smaller than the coherence length of interstellar turbulence. The Larmor radius for this energy range is estimated to be 
\begin{equation}
 r_{L,i}=\frac{E}{Z_ieB_{\rm ISM}}\simeq 1.1\times10^{-6} \mathcal R_1B_{\rm ISM,-5}^{-1} \rm\, pc,
\end{equation}
where $B_{\rm ISM}$ is the magnetic field in ISM and $\mathcal R_1=\mathcal R/(10\rm\,GV)$. This can be smaller than the typical coherence length of the interstellar turbulence, $\lambda_c \sim 10-100$ pc \citep{Han08a}. Hence, we can use the same value of $\delta$ for $\mathcal R\lesssim10^8$ GV. Also, this estimate implies that the motion of the CRs are likely diffusive rather than ballistic.

\subsection{Intensity and composition}
\begin{table}
\begin{center}
\caption{Model parameters used in this work. \label{tab:models}}
\begin{tabular}{|c|ccccc|}
\hline
 model & $M_{\rm ej}\rm \,[\msun]$ & $V_{\rm ej}\,[c] $ & $\epsilon_B$ & $B_{\rm ISM}$ [$\rm\mu$G] & $\epsilon_{\rm CR}$ \\
\hline
 A & 0.03  &  0.25 & $10^{-3}$ & --- & 0.2 \\
 B & 0.05  &  0.3 & --- & 8 &  0.08 \\
\hline
\end{tabular}
\end{center}
\end{table}

  \begin{figure}
   \begin{center}
    \includegraphics[width=\linewidth,pagebox=cropbox,clip]{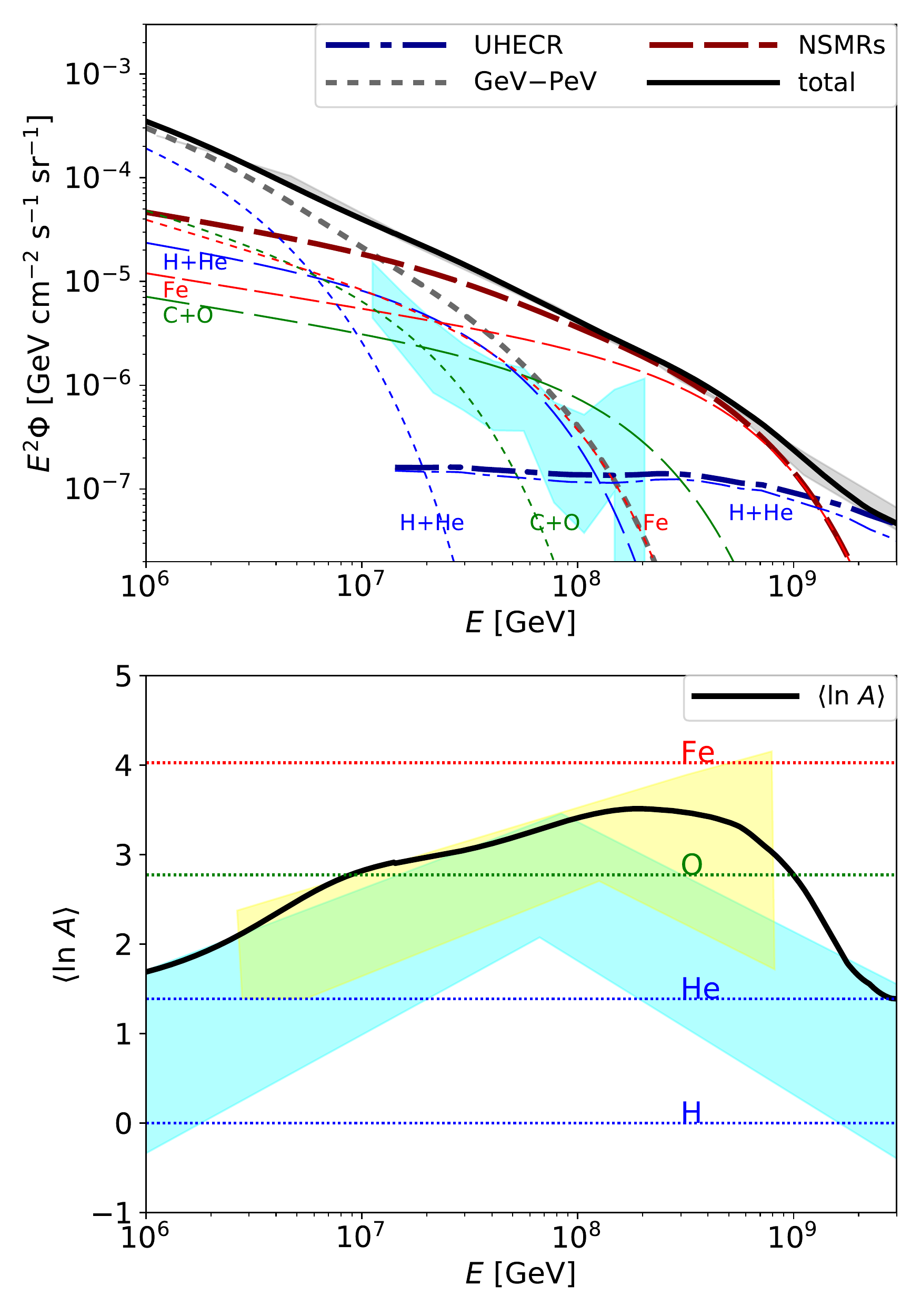}
    \caption{Upper panel: Comparison of the CR spectrum in the NSMR model to the experimental data. The thick-solid line is the total flux estimated by our model. The thick-dashed line represents the NSMRs (our work). The thick-dotted and thick-dot-dashed lines indicate the GeV--PeV and the UHECR components, respectively. See the text for the details of these components. The color-thin lines show the spectrum for each element group: H+He (blue), CNO (green), and Fe (red). The color-dashed, color-dotted, and color-dot-dashed lines are for the NSMR, the GeV--PeV, and the UHECR components, respectively. The experimental data for the total flux are taken from \citet{VIT17a} and \citet{TA18a}, which are written in gray band. The flux data for the light elements (H+He) shown in the cyan region are taken from \citet{KASCADE13a}.  Lower panel: $\langle \ln A \rangle$ as a function of energy. The experimental data are taken from \citet{KU12a} (cyan region) and \citet{Gai16a} (yellow region). The thick-solid line is the model calculation. The parameters are set to be $n_{\rm ISM}=0.1\rm\, cm^{-3}$, $\delta=1/3$, $\rho_{\rm MW}=1.5\times10^{-4}\rm\,yr^{-1}$, $M_{\rm gas}=10^{10}\rm\,\msun$, $s_{\rm inj}\approx s_{\rm esc}=2.0$, and the other parameters are tabulated in Table \ref{tab:models}. The results for model A and B are almost indistinguishable.}
    \label{fig:result}
   \end{center}
  \end{figure}

We approximate the total CR production energy per merger to be $\mathcal E_{\rm cr}\approx \epsilon_{\rm cr} M_{\rm ej}V_{\rm ej}^2/2$, where $\epsilon_{\rm cr}$ is the production efficiency of CRs. We assume that the spectrum of CRs escaping from the NSMRs  is a power-law with exponential cutoff: $dN_i/dE \propto E^{-s_{\rm esc}}\exp(-E/E_{i,\rm max})$.  Then, the differential CR production rate by the NSMRs for species $i$ is approximated to be
\begin{equation}
(EQ_{E,\rm inj})_i\approx \frac{f_i \mathcal E_{\rm cr}\rho_{\rm MW} }{\ln\left(E_{p,\rm max}/E_{p,\rm min}\right)}\exp\left(-\frac{E}{E_{i,\rm max}}\right),
\end{equation}
where $f_i$ is the abundance ratio shown in Section \ref{sec:composition} and we set $s_{\rm inj}\approx s_{\rm esc}= 2$. The normalization factor, $\ln(E_{p,\rm max}/E_{p,\rm min})$, is estimated by using the maximum and minimum energy for protons, and $E_{p,\rm min}$ is set to 1 GeV.

We use the grammage to estimate the spectrum in the CR halo. The Boron-to-Carbon ratio (B/C) obtained by the recent experiments \citep{PAMELA14a,AMS16a} enables us to estimate the grammage traversed by CRs to be~\citep[e.g.,][]{BKW13a}
\begin{equation}
 X_{\rm esc} \simeq 8.7 \mathcal R_1^{-\delta} \,\rm g\, cm^{-2}.\label{eq:Xesc}
\end{equation}
We use $\delta=0.46$ for $\mathcal R<250$ GV and $\delta=1/3$ for $\mathcal R\ge250$ GV \citep{MF18a}. The escaping rate of CRs from the CR halo is written as $E U_{E} c M_{\rm gas}/X_{\rm esc}$, where $U_{E}$ is the differential energy density of the CRs of species $i$ and $M_{\rm gas}\sim 10^{10}\rm\, \msun$ is the gas mass inside the Milky Way galaxy.
Equating the injection rate and the escape rate, we obtain~\citep[e.g.,][]{MF18a}
\begin{equation}
(E^2 \Phi)_i\approx \frac{(EQ_{E,\rm inj})_i X_{\rm esc}}{4\pi M_{\rm gas}}\propto E^{-\delta}\exp\left(-\frac{E}{E_{i,\rm max}}\right).
\end{equation}
Note that the normalization of the intensity is independent of the escape time, $T_{\rm esc}$, that has larger uncertainty depending on propagation models.

The resulting spectrum is shown in the upper panel of Figure \ref{fig:result}, whose  parameter set is summarized in Table \ref{tab:models} as model A (see Section \ref{sec:discussion} for model B. The results are almost identical to those for model A). We also plot two additional components, the GeV--PeV and UHECR components, which account for the regions below the knee and above the ankle, respectively. For the GeV--PeV component, the spectral shape is assumed to be a power-law and an exponential cutoff with the spectral index of $-2.6$ and the cutoff energy of $2\times10^{15}Z_i$ eV. The abundance ratio for this component is set to be the same as that for the observed CRs at 1 TeV: $(f_p,\, f_{\rm He},\, f_{\rm C},\, f_{\rm O},\, f_{\rm Ne},\, f_{\rm Si},\, f_{\rm Fe})\simeq$  (0.43,  0.28,  0.052,  0.077,  0.023, 0.039, 0.10) \citep[see, e.g., ][]{WBM98a,Hor03a}. We set the normalization of the GeV--PeV component so that it fits the data. For the UHECR component, we use the model by \citet{FM18a} that fits the observed UHECR data above $2\times10^{18}$ eV as well as the ankle feature around $10^{17}$ eV for the CR proton and helium flux. In the range of our interest, the UHECR component is light-element dominant.

We can see that the overall spectrum is well fitted by the three components: the experimental data (gray band) is almost completely overlapped with the model curve (thick-solid line). The NSMR component is dominant for the energy range $2\times10^{16}$ eV -- $10^{18}$ eV. This causes a slight hardening for the total flux at $E\sim 2\times10^{16}$ eV, which is consistent with the recent experiments \citep{IceTop13a,TA18a}.

The abundance ratio of the CRs from the NSMRs on Earth is $(f_p,\, f_{\rm He},\, f_{\rm C},\, f_{\rm O},\, f_{\rm Ne},\, f_{\rm Si},\, f_{\rm Fe}) \simeq (0.10,\,0.41,\,0.028,\,0.13,\,0.037,\,0.043,\,0.26)$. This helium abundance is higher than that for the GeV-PeV component.
This arises from the injection prescription with the assumption of a singly ionized plasma (see Section \ref{sec:composition}). The ionization degree in the ISM could be higher, which suppresses the CR nuclei production compare to CR protons. Also, majority of heavy elements exist in dust grains, which enhances the injection efficiency of the heavy elements \citep{EDM97a}. These effects can decrease the helium abundance. Detailed modeling including these effects are beyond the scope of this paper.  

Our model is also in good agreement with the spectrum for the light elements (H + He), which show a hardening around $10^{17}$ eV \citep{KASCADE13a,16BCFa}. The light elements from the NSMRs match the observed spectrum below the hardening, and the UHECR component accounts for the energy range above the hardening. The contribution from the GeV--PeV component is negligible there because of the lower cutoff energy. The bottom panel shows the average mass number, $\langle \ln A\rangle$. This is also consistent with the experimental results, although they have large uncertainty.

\section{Discussion}\label{sec:discussion}
We use a simple assumption about the magnetic field amplification using a constant $\epsilon_B$ parameter, which is widely used to discuss the afterglow emission of gamma-ray bursts and NSMRs \citep{Mes06a,KZ15a}. Recent particle simulations \citep{CS14b,MCM18a} suggest that the magnetic field amplification at non-relativistic shocks is represented as 
\begin{equation}
 B \sim \sqrt{0.5 \mathcal M_A} B_{\rm ISM}\simeq 0.25 V_{\rm ini,-0.6}^{1/2}n_{\rm ISM,-1}^{1/4} B_{\rm ISM,-5}^{1/2}\rm\, mG,
\end{equation}
where $\mathcal M_A=V_{\rm ej}/V_A$ is the Alfven mach number and $V_A=B_{\rm ISM}/\sqrt{4\pi \mu m_p n_{\rm ISM}}$ is the Alfven velocity at the upstream. Even using this formalism of magnetic field amplification, our model can explain the CRs around the second knee with an optimistic set of parameters tabulated in Table \ref{tab:models} as model B. The resulting spectrum and composition are almost identical to those shown in Figure \ref{fig:result}.

Although we set the spectrum index at the source to $s_{\rm inj}=2$, it could be softer. It is widely believed that Galactic SNRs accelerate the CRs below the knee, where the spectral index at the Earth is $\gamma\approx2.7-2.8$. This results in the spectral index at the sources, $s_{\rm esc}\approx2.4$~\citep{MF18a}. 
The spectral softening can be caused by a number of reasons~\citep[e.g.,][]{OI11a}. 
A softer escape spectrum requires a higher CR production rate to achieve the observed CR intensity. Since we use a fairly high value of $\epsilon_{\rm CR}$ for model A, this might cause some tension between our model and experimental data.

When BNSs merge, faster ejecta can be dynamically produced due to the shock formed by the merger. This dynamical ejecta can be faster ($\beta_{\rm ej}=V_{\rm ej}/c\lesssim 0.8$) and lighter ($M_{\rm ej}\sim 0.01\msun$). According to the afterglow observations of GW170817, the kinetic energy distribution of the dynamical ejecta can be $\mathcal E_k(>\Gamma_{\rm ej}\beta_{\rm ej})\propto (\Gamma_{\rm ej}\beta_{\rm ej})^{-5}$ \citep{MNH18a}. With this steep profile, the slower shell accelerates CRs to higher energy than the faster shell, because the faster shell is decelerated too quickly to accelerate CRs. Thus, we can neglect the CRs produced by the fast tail of the dynamical ejecta.

CRs are produced also at reverse shocks of the NSMRs. Since the merger ejecta consist of $r$-process elements, these CRs should be heavier than iron. However, the CR experiments around GeV energy suggest that there is no strong enhancement of $r$-process elements \citep{BGG89a,DTO12a}, which limits the CR production efficiency at the reverse shocks of NSMRs to be lower than $3\times10^{-5}$ \citep{KI16a}. In our model, the CRs produced at the reverse shock should be confined inside the ejecta, and are expected to lose energies by adiabatic expansion so that the $r$-process elements do not have to contribute to the observed CRs.

The central remnant object left after a BNS merger could be a magnetar. In this case, it produces the magnetar wind with the total energy $\mathcal E_w\sim10^{52}$ erg if the rotation period is $\sim 1$ ms \citep[e.g.,][]{YZG13a,MP14a,MTF18a}. 
The energy and mass of the wind are deposited to the ejecta of the merger, so the ejecta of NSMRs can be faster and more massive than those we assumed in this paper, leading to mildly relativistic ejecta. This kind of energetic ejecta can produce higher-energy CRs, so the production of UHECRs is possible. If one third of BNS mergers have magnetars left, and 30 \% of ejecta energy is spent to produce CRs, the luminosity density for UHECR production at $E\sim 10^{19}$ eV is roughly estimated to be $\mathcal C \epsilon_{\rm cr}\mathcal E_w \rho_{\rm mer}/3\sim 10^{44}\rm\, erg\, Mpc^{-3}\,yr^{-1}$, where $\mathcal C\sim1/15$ is the bolometric correction factor. This value is consistent with the required luminosity density of observed UHECRs \citep[e.g.,][]{KBW09a,MT09a}.  Note that the central remnant of GW170817 should be a black hole, because we do not observe the luminous X rays and high-energy gamma rays in later time from GW170817 (see \citealt{MTF18a} for related discussions). The magnetar model will be examined by the counterpart searches of GWs \citep[see][and references therein]{LIGO17d} and transient searches in the radio band \citep{VLASS13a,GMRT16a,ThunderKAT17a}.

The heavy elements might be disintegrated by interaction with photons during the propagation in ISM. For iron nuclei with energy $\sim 10^{18}$ eV, the target photons for the giant dipole resonance is $E_\gamma\sim$ 1 eV. The energy density of infrared radiation field is $\sim 1\rm\, eV\, cm^{-3}$, leading to the photon number density $n_\gamma\sim 1\rm\, cm^{-3}$. The photodisintegration timescale is then $t_{A\gamma}\sim (n_\gamma \sigma_{A\gamma}c)^{-1}\sim$ 10 Myr, where $\sigma_{A\gamma}\sim 10^{-25} \rm\,cm$ is the giant dipole resonant cross-section for iron nuclei. Although this is sufficiently longer than the escape time of CRs of $E_{\rm Fe}\sim10^{18}$ eV, the radiation density is much higher in the Galactic Center, where the CR nuclei would be destroyed. The detailed simulation including this effect is beyond the scope of this paper. Note that since the CRs are produced during the Sedov phase ($t_{\rm dec}\gtrsim 10$ years), we safely neglect the disintegration at the NSMR. Also, the spallation by interaction with ISM is not important in this energy range.

The NSMRs can be located above the Galactic disk because of the natal kick. The spacial distribution of NSMRs is likely to be similar to the distribution of milisecond pulsars \citep{Lor13a} and X-ray binaries \citep{RIN17a}. The scale height of the CR halo, $H_h$, is often set to be 4 kpc -- 6 kpc \citep[e.g.,][]{SMP07a}, which is larger than the distributions of these objects. Hence, the Galactic BNS mergers are likely to occur inside this CR halo. 
Then, the NSMRs intermittently inject the CRs at random positions in the CR halo, which results in an inhomogeneous CR distribution. This causes the dipole anisotropy.
The dipole anisotropy of CRs from the isotropically distributed sources in the disk is roughly estimated to be \citep[e.g.,][]{BA12a}
\begin{equation}
 a \sim \frac{D}{c H_h}\sim 1.7\times10^{-4} \mathcal R_1^\delta  H_{h,22.2},
\end{equation}
where $D$ is the diffusion coefficient and we set $H_h\sim 5$ kpc and $D\sim 8\times10^{28}\mathcal R_1^\delta \rm\,cm^2 s^{-1}$. For $10^{16}{\rm\,eV} < E < 10^{18}\,\rm eV$, CR arrival direction is almost isotropic and upper limits for the amplitude of dipole anisotropy are $\sim0.01$ \citep{KASCADE04a,PAO-ICRC15}. Our rough estimate results in $a \sim$ a few percents for $\mathcal R\sim 10^8$ GV, which might have a tension with the observations. However, NSMRs can be located above the Galactic disk, so the estimate above is not appropriate in our situation. The realistic value of $a$ should be calculated by solving the anisotropic diffusion equations, 
appropriately taking into account magnetic field geometry and intermittent and inhomogeneous CR sources \citep[see e.g.,][for lower energy CRs]{BA12b,PE13a,MF15a}.

Mergers of neutron star-black hole (NS-BH) binaries are also expected to produce energetic outflows 
\citep{Ros05a,EPS12a,FDD14a,KSK15a,KIO15a}, which is likely to produce CRs as is the case with BNS mergers. The numerical simulations indicate that NS-BH mergers can produce more energetic outflows than those by BNS mergers \citep{JBA15a,WFM16a}. Although the NS-BH merger rate and its ejecta energy is more uncertain at present, the planned GW experiments and counterpart searches will reveal the physical quantities in the system near future.

\section{Summary}\label{sec:summary}
We have investigated the CR production due to Galactic NSMRs, based on the parameters estimated from the observations of GW170817. 
Assuming efficient amplification of the magnetic field at forward shocks, we have found that the NSMRs can accelerate protons and iron nuclei up to $2\times10^{16}$ eV and $5\times10^{17}$ eV, respectively. The event rate of BNS mergers is high enough to provide CRs before they escape from the Galaxy. 
Using a simple model that takes into account the escape process during the Sedov phase and the enhancement of heavy elements in the CR injection, we have calculated the spectrum and composition of the CRs on Earth. 
Together with the GeV-PeV CR and UHECR components, our model can be consistent with the observed CR data. The NSMR component may give a significant contribution to the CR flux at energies of $10^{16}$~eV $\lesssim E \lesssim 10^{18}$~eV, and the other components should account for the CRs below the knee and above the ankle. Our model could also match the spectrum for the light elements (H and He) around $10^{16}$~eV $\lesssim E \lesssim 3\times 10^{17}$~eV.

\acknowledgments
S.S.K thanks Susumu Inoue for useful discussion at the ASJ annual meeting on March 2018. We are grateful to Kunihito Ioka for helpful comments and Ke Fang for providing the data. This work is supported by JSPS Oversea Research Fellowship, the IGC post-doctoral fellowship program (S.S.K.), Alfred P. Sloan Foundation, NSF Grant No. PHY-1620777 (K.M.), and NASA NNX13AH50G (P.M.).  While we were finalizing this project, we became aware of a related but independent work by Rodrigues et al. (arXiv:1806.01624). They focus on extragalactic BNS mergers and CR protons around the ankle, whereas we concentrate on the Galactic NSMRs and CR irons around and beyond the second knee.

\bibliographystyle{hapj}
\bibliography{ssk}


\listofchanges

\end{document}